\documentclass[onecolumn,showpacs]{revtex4}
\usepackage{epsfig}
\usepackage{psfig}
\usepackage{graphicx}%
\usepackage{dcolumn}
\usepackage{amsmath}

\makeatletter
\def\btt#1{\texttt{\@backslashchar#1}}%
\DeclareRobustCommand\bblash{\btt{\@backslashchar}}%
\makeatother

\begin{document}

\title{VISCOUS GENERALIZED CHAPLYGIN GAS}
\
\author{XIANG-HUA ZHAI}

\author{YOU-DONG XU}

\author{XIN-ZHOU LI}\email{kychz@shtu.edu.cn}

\affiliation{Shanghai United Center for Astrophysics(SUCA), Shanghai
Normal University, 100 Guilin Road, Shanghai 200234, China
}%
\date{\today}

\begin{abstract}
 Viscous GCG(generalized Chaplygin gas) cosmology is discussed,
assuming that there is  bulk viscosity in the linear barotropic
fluid and GCG. The dynamical analysis indicates that the phase
$w_g=-1+\sqrt{3}\gamma\kappa\tau_{g}/(\gamma-\sqrt{3}\kappa\tau_{\gamma})$
is a dynamical attractor and the equation of state of GCG approaches
it from either $w_g>-1$ or $w_g<-1$ depending on the choice of its
initial cosmic density parameter and the ratio of pressure to
critical energy density. Obviously, the equation of state $w_g$ can
cross the boundary $w_g=-1$. Also, from the point of view of
dynamics, the parameters of viscous GCG should be in the range of
$\gamma>\sqrt{3}\kappa\tau_{\gamma}/(1-\sqrt{3}\kappa\tau_{g})$ and
$0<\alpha<1+\sqrt{3}
\kappa\gamma\tau_{g}/(\gamma-\sqrt{3}\kappa\tau_{\gamma}-\sqrt{3}\kappa\gamma\tau_{g})$.

\end{abstract}


\maketitle

\vspace{0.4cm}\noindent\textbf{1. Introduction} \vspace{0.4cm}

CMB anisotropy, supernovae and galaxies clustering strongly indicate
that our universe is spatially flat, with two thirds of the energy
content consisting of dark energy, a substance with negative
pressure which can make the universe expand in an accelerating
fashion. Present observation data constrain the range of the
equation of state of dark energy as $-1.38<w<-0.82$
\cite{melchiorri}, which indicates the possibility of dark energy
with $w<-1$. If this is so, the universe can have some strange
properties such as a future finite singularity which has been dubbed
Big Rip\cite{caldwell} or Big Smash\cite{mcinnes}. Proposed
candidates for dark energy include the cosmological constant,
quintessence with a single field\cite{Peebles} or with $N$ coupled
fields\cite{li}, phantom field with canonical\cite{well} or
Born-Infeld type Lagrangian\cite{hao3}, $k$-essence\cite{armendariz}
and generalized Chaplygin gas(GCG)\cite{GCG} which is based on the
Chaplygin gas\cite{chgas}. In particular, we have extended the
equation of state of GCG to the regime $w<-1$ regime and shown that
the GCG parameter is constrained by the dynamics to lie in the range
$0<\alpha<1$ \cite{hao}.

On the other hand, the role of dissipative processes has been
conscientiously studied\cite{barrow}. Dissipative effects, including
both bulk and shear viscosity, play a very important role in the
evolution of the universe. The viscosity theory of relativistic
fluids was first suggested by Eckart\cite{eckart} and Landau and
Lifshitz\cite{landau}, who considered only first-order deviation
from equilibrium, which lead to parabolic differential equations and
hence to an infinite speed of propagation for heat flow and
viscosity, in contradiction with the principle of causality. The
relativistic second-order theory was founded by Israel\cite{israel}
and developed by Israel and Stewart\cite{stewart}, and has also been
used in the evolution of the early universe\cite{harko}. However,
the character of the evolution equation is very complicated in the
framework of the full causal theory. Therefore, the conventional
theory\cite{landau} is still applied to phenomena which are
quasi-stationary, i.e., slowly varying on space and time scales
characterized by the mean free path and the mean collision time. In
the case of isotropic and homogeneous cosmologies, the dissipative
process can be modelled as a bulk viscosity $\zeta$ within a
thermodynamical approach. The shear viscosity $\eta$ will be
neglected, which is consistent with the usual practice\cite{brevik}.
The bulk viscosity introduces dissipation by only redefining the
effective pressure, $p_{eff}$, according to $p_{eff}=p-3\zeta H$
where $\zeta$ is the bulk viscosity coefficient and $H$ is the
Hubble parameter. The condition $\zeta>0$ assures a positive entropy
production in conformity with the second law of
thermodynamics\cite{zimdahl}. We are interested in the case
$\zeta=\sqrt{3}\kappa^{-1}\tau H$, where $\tau$ is a constant. This
assumption implicates that $\zeta$ is directly proportional to the
divergence of the cosmic fluid's velocity vector. Therefore, it is
physically natural, and has been considered previously in an
astrophysical context\cite{gron}.

In the present paper, we consider a viscous GCG cosmological model
for the expanding universe, assuming that there is bulk viscosity in
the linear barotropic fluid and GCG. The dynamical analysis
indicates that the phase
$w_g=-1+\sqrt{3}\gamma\kappa\tau_{g}/(\gamma-\sqrt{3}\kappa\tau_{\gamma})$
is a dynamical attractor and the equation of state of GCG approaches
it from either $w_g>-1$ or $w_g<-1$ depending on the choice of its
initial cosmic density parameter and the ratio of pressure to
critical energy density, where we assume $\sqrt{3}\kappa\tau_g$ and
$\sqrt{3}\kappa\tau_{\gamma}$ are small compared to $1$ . If indeed
dark energy with $w<-1$ is within the regime of possibilities, then
it would seem inevitable to inquire about a transition from $w<-1$
to $w>-1$, although a generalized scalar field cannot safely
traverse $w=-1$. It is clear that $w$ can cross the boundary
$w_g=-1$ in the viscous GCG cosmology. Also, we have used
cosmological dynamics to place constraints on some of the
parameters.

\vspace{0.4cm} \noindent\textbf{2. Viscous GCG}
 \vspace{0.4cm}

GCG has a very simple equation of state,
$p_{g}=-M^{4(\alpha+1)}/\rho_{g}^{\alpha}$, which yields an
analytically solvable cosmological dynamics if the universe is GCG
dominated. Another advantage of introducing GCG is to unify dark
energy and dark matter into one equation of state, also known as
quartessence\cite{reis}. However, detailed numerical analysis turns
out to disfavor the dark matter modelled by the GCG equation of
state\cite{sandvik}. But no observation has so far ruled out the
possibility of GCG as dark energy. Therefore, it is quite possible
that our universe contains a dark energy component modelled by the
GCG as well as another linear barotropic fluid component with the
equation of state $p_{\gamma}=(\gamma-1)\rho_{\gamma}$. However, in
this section, we focus first on the viscous GCG system, as the
cosmological dynamics are analytically solvable if GCG is dominant .
In the flat FRW universe, the field equations are

 \begin{eqnarray}\label{friedmann}
 H^2&=&(\frac{\dot{a}}{a})^2=\frac{\kappa^2}3\rho_g\\
 \dot{H}+H^2&=&\frac{\ddot{a}}a=-\frac{\kappa^2}6(\rho_g+3p_g-9H\zeta)
 \end{eqnarray}

\noindent  The conservation equation is

\begin{equation}\label{conserv}
\dot{\rho_g}+3H(\rho_g+p_g-3H\zeta)=0
\end{equation}

\noindent Using the GCG equation of state, we obtain

\begin{equation}\label{relat}
\frac a 3
\frac{d\rho_g}{da}+\rho_g-\frac{M^{4(\alpha+1)}}{\rho_g^{\alpha}}-\sqrt{3}\kappa\rho_g^{\frac12}\zeta=0
\end{equation}

\noindent We shall be interested in the evolution of the late
universe, from $t=t_0$ onwards. From Eq.(4), we have

\begin{equation}\label{scale}
a=a_0\Big(\exp\Big[\int_{\rho_0}^{\rho_g}\frac{\rho^{\alpha}d\rho}
{M^{4(\alpha+1)}+\sqrt{3}\kappa\rho^{\alpha+\frac12}\zeta(\rho)-\rho^{\alpha+1}}\Big]\Big)^{\frac13}
\end{equation}

\noindent which is the general relation between the cosmological
scale factor $a$ and the energy density $\rho_g$.

In what follows we investigate two different solvable cases. If we
choose $\zeta=\tau_g\sqrt{\rho}$, the energy density is given by

\widetext
\begin{equation}\label{density}
\rho_g=\Big[\frac{M^{4(\alpha+1)}}{1-\sqrt{3}\kappa\tau_g}
+\Big(\rho_0^{\alpha+1}-\frac{M^{4(\alpha+1)}}{1-\sqrt{3}
\kappa\tau_g}\Big)\Big(\frac{a_0}a\Big)^{3(\alpha+1)(1-\sqrt{3}\kappa\tau_g)}\Big]^{\frac1{\alpha+1}}
\end{equation}
\endwidetext

\noindent Obviously, $\mid p_g\mid \rightarrow\infty$ or $0$ will
not happen when $a\rightarrow a_s\neq0$,corresponding to no future
singularities. As $a\rightarrow\infty$, we have $w_g\rightarrow
-(1-\sqrt{3}\kappa\tau_g)>-1$.

As a second case, if we take $\zeta=\tau_g\rho^{\alpha+\frac32}$
 then Eq.(4)
reads
\begin{equation}\label{case2}
\frac{a}{3(\alpha+1)}\frac{d}{da}(\rho_g^{\alpha+1})=-\rho_g^{\alpha+1}
+M^{4(\alpha+1)}+\sqrt{3}\kappa\tau_g\rho_g^{2(\alpha+1)}
\end{equation}

\noindent and if $\tau_g<(4\sqrt{3}\kappa M^{4(\alpha+1)})^{-1}$,
it has two stationary points at
 $\rho_g=\Big(\frac{1-\Delta}{2\sqrt{3}\kappa\tau_g}\Big)^{\frac
1{\alpha+1}}\equiv \rho_1$ and
$\rho_g=\Big(\frac{1+\Delta}{2\sqrt{3}\kappa\tau_g}\Big)^{\frac
1{\alpha+1}} \equiv \rho_2$ where $\Delta=(1-4\sqrt{3}\kappa\tau_g
M^{4(\alpha+1)})^{\frac 12}$. The first stationary point is an
attractor, and corresponds to $a \rightarrow\infty$. The second
stationary point is unstable, and corresponds to $a=0$. If
$\rho_0<\rho_1$ or $\rho_1<\rho_0<\rho_2$ then the model evolves As
$a\rightarrow\infty$, $\rho_g\rightarrow\rho_1$ and $w_g\rightarrow
-\frac{2\sqrt{3}\kappa\tau_g M^{4(\alpha+1)}}{1-\Delta}>-1$.
However, if $\rho_0>\rho_2$ then as $a\rightarrow
a_0\Big(\frac{\rho_0^{\alpha+1}-\rho_1^{\alpha+1}}{\rho_0^{\alpha+1}-\rho_2^{\alpha+1}}\Big)
^{\frac{1}{3\Delta(\alpha+1)}}, \rho_g\rightarrow\infty$ and
$p_g\rightarrow\infty$ so the model encounters a Type III
singularity(where the pressure and density diverge in finite proper
time but the metric remains non-singular).

 \vspace{0.4cm} \noindent\textbf{3. Autonomous system}
 \vspace{0.4cm}

A general study of the phase space system of quintessence  and
phantom in FRW universe has been given in Refs.\cite{haonew, li2}.
For the viscous GCG cosmological dynamical system, the
corresponding equation of motion and Einstein equations can be
written as

\begin{eqnarray}\label{sys}
\dot{H}&=&-\frac{\kappa^2}{2}(\rho_\gamma+p_\gamma-3\zeta_{\gamma}H+\rho_{g}+p_g-3\zeta_{g}H)\\
\dot{\rho_\gamma}&=&-3H(\rho_\gamma+p_\gamma-3\zeta_{\gamma}H)\\
\dot{\rho_{g}}&=&-3H(\rho_{g}+p_{g}-3\zeta_{g}H)\\
H^2&=&\frac{\kappa^2}{3}(\rho_\gamma+\rho_{g})
\end{eqnarray}

\noindent where $\rho_{\gamma}$ is the density of fluid with a
barotropic equation of state $p_{\gamma}=(\gamma-1)\rho_{\gamma}$,
and $0\leq\gamma\leq 2$ is a constant that relates to the equation
of state by $w_{\gamma}=\gamma-1$; $\rho_g$ and $p_g$ are the
energy density and pressure of GCG, respectively. The overdot
represents the derivative with respect to cosmic time $t$. We
choose $\zeta_g=\tau_g\sqrt{\rho}=\sqrt{3}\kappa^{-1}\tau_g H$ and
$\zeta_{\gamma}=\tau_{\gamma}\sqrt{\rho}=\sqrt{3}\kappa^{-1}\tau_{\gamma}H$.

To analyze the dynamical system, we rewrite the equations with the
following dimensionless variables:

\begin{eqnarray}\label{vari}
x&=&\frac{\kappa^2\rho_{g}}{3H^2}\nonumber\\
y&=&\frac{\kappa^2p_{g}}{3H^2}\nonumber\\
N&=&\ln a
\end{eqnarray}

\noindent The dynamical system can be reduced to
 \widetext
\begin{eqnarray}\label{auto}
\frac{dx}{dN}&=&-3(x+y)+3\sqrt{3}\kappa\tau_g+3x[\gamma(1-x)+x+y]
-3\sqrt{3}\kappa(\tau_g+\tau_{\gamma})x\\
\frac{dy}{dN}&=&3\alpha(y+\frac{y^2}x)-3\sqrt{3}\alpha\kappa\tau_g\frac{y}
{x}+3y[\gamma(1-x)+x+y]-3\sqrt{3}\kappa(\tau_g+\tau_{\gamma})y
\end{eqnarray}
\endwidetext
\noindent From Eq.(11), we have

\begin{equation}\label{constraint}
\Omega_{g}+\Omega_{\gamma}=1
\end{equation}

\noindent where $\Omega_{g}\equiv x$ and $\Omega_{\gamma}\equiv
\frac{\kappa^2\rho_{\gamma}}{3H^2}$ are the cosmic density
parameters for GCG and linear barotropic fluid, respectively. The
equation of state can be expressed in terms of the new variables as

\begin{equation}\label{equaofstate}
 w_{g}=\frac{p_{g}}{\rho_{g}}=\frac{y}{x}
\end{equation}

\noindent and the sound speed is

\begin{eqnarray}\label{sound}
c_s^2= -\alpha\frac{y}{x}
\end{eqnarray}

In the following we need to consider the two cases of $\gamma=1$
and $\gamma\neq 1$. In the case of $\gamma\neq 1$, the critical
points of the system are

\begin{eqnarray}\label{crit}
(x^{(1)},
y^{(1)})&=&\Big(\frac{\gamma-1-\sqrt{3}\kappa\tau_g-\sqrt{3}\kappa\tau_{\gamma}-\Sigma}{2(\gamma-1)},
0\Big)\\
(x^{(2)},
y^{(2)})&=&\Big(\frac{\gamma-1-\sqrt{3}\kappa\tau_g-\sqrt{3}\kappa\tau_{\gamma}+\Sigma}{2(\gamma-1)},
0\Big)
\end{eqnarray}

\noindent and
\begin{equation}
(x^{(3)},
y^{(3)})=\Big(1-\frac{\sqrt{3}\kappa\tau_{\gamma}}{\gamma},
-1+\sqrt{3}\kappa\tau_g+\frac{\sqrt{3}\kappa\tau_{\gamma}}{\gamma}\Big)
\end{equation}

\noindent which correspond to the linear barotropic fluid-dominated
phase, GCG matter-dominated phase and GCG vacuum-energy-dominated
phase, respectively. Here
\begin{equation}
\Sigma=[4\sqrt{3}(\gamma-1)\kappa\tau_g+(\gamma-1-\sqrt{3}\kappa\tau_g-\sqrt{3}\kappa\tau_{\gamma})^2]^{\frac12}
\end{equation}

\noindent If we linearize the system near the critical points
$(x^{(i)}, y^{(i)}), i=1,2,3$ and then translate the system to the
origin, we can readily write the first order perturbation equation
as

\begin{equation}\label{perturb}
\frac{dU}{dN}=A^{(i)}U
\end{equation}

\noindent where $U$ is a 2-column vector consisting of the
perturbations of $x$ and $y$. $A^{(i)}$ is a $2\times2$ matrix for
the critical point $(x^{(i)}, y^{(i)})$. The stability of the
critical points is determined by the eigenvalues of the matrix
$A^{(i)}$ at the critical point $(x^{(i)}, y^{(i)})$. For the
point $(x^{(1)}, y^{(1)})$, the two eigenvalues are

\begin{eqnarray}\label{eigen1}
\lambda_{1}^{(1)}&=&3\Sigma\\\nonumber \lambda_{2}^{(1)}&=&
\frac{3(1+\alpha)(\gamma-1+\sqrt{3}\kappa\tau_g-2\sqrt{3}\gamma\kappa\tau_g-
\sqrt{3}\kappa\tau_{\gamma}-\Sigma)}{\gamma-1-\sqrt{3}\kappa\tau_g-\sqrt{3}\kappa\tau_{\gamma}-\Sigma}
\end{eqnarray}

\noindent Obviously, the linear barotropic fluid-dominated phase is
unstable, and so evolves to GCG-dominated phase. For the point
$(x^{(2)}, y^{(2)})$ the two eigenvalues are

\begin{eqnarray}\label{eigen1}
\lambda_{1}^{(2)}&=&-3\Sigma\\\nonumber \lambda_{2}^{(2)}&=&
\frac{3(1+\alpha)(\gamma-1+\sqrt{3}\kappa\tau_g-2\sqrt{3}\gamma\kappa\tau_g-
\sqrt{3}\kappa\tau_{\gamma}+\Sigma)}{\gamma-1-\sqrt{3}\kappa\tau_g-\sqrt{3}\kappa\tau_{\gamma}+\Sigma}
\end{eqnarray}

\noindent It can be shown that $\lambda_2^(2)>0$, and so the point
is stable if $\alpha
> -1$ and
$\gamma<\frac{\sqrt{3}\kappa\tau_{\gamma}}{1-\sqrt{3}\kappa\tau_g}$.
However, since we want the GCG to first behave as matter and then
evolve to behave like dark energy, it will not be appropriate if
$(x^{(2)}, y^{(2)})$ corresponds to a stable attractor phase. In
other words
$\gamma<\frac{\sqrt{3}\kappa\tau_{\gamma}}{1-\sqrt{3}\kappa\tau_g}$
should not be considered in the real models. For the critical point
$(x^{(3)}, y^{(3)})$, the corresponding eigenvalues of matrix
$A^{(3)}$ are

\begin{eqnarray}\label{eigen3}
\lambda_1^{(3)}&=&-\frac
{3}{2(\gamma-\sqrt{3}\kappa\tau_{\gamma})}(\Theta+\Xi)\\\nonumber
\lambda_2^{(3)}&=&-\frac
{3}{2(\gamma-\sqrt{3}\kappa\tau_{\gamma})}(\Theta-\Xi)
\end{eqnarray}

\noindent where

\widetext
\begin{eqnarray}
\Theta&=&\gamma^2+[(1+\alpha)(1-\sqrt{3}\kappa\tau_g)
-2\sqrt{3}\kappa\tau_{\gamma}]\gamma-(1+\alpha)\sqrt{3}\kappa\tau_{\gamma}+3\kappa^2(\tau_g\tau_{\gamma}+\tau_{\gamma}^2)\\\nonumber
\Xi&=&\Big(-4(1+\alpha)(\gamma-\sqrt{3}\kappa\tau_{\gamma})^2
[\gamma(1-\sqrt{3}\kappa\tau_g)-\sqrt{3}\kappa\tau_{\gamma}]\\
&+&\big[\gamma^2+[(1+\alpha)(1-\sqrt{3}\kappa\tau_g)
-2\sqrt{3}\kappa\tau_{\gamma}]\gamma-(1+\alpha)\sqrt{3}\kappa\tau_{\gamma}+3\kappa^2(\tau_g\tau_{\gamma}+\tau_{\gamma}^2)
\big]^2\Big)^{\frac {1} {2}}\nonumber
\end{eqnarray}
\endwidetext

\noindent Note that
$\lambda_1^{(3)}\lambda_2^{(3)}=9(\alpha+1)(\gamma-\sqrt{3}\gamma\kappa\tau
_g-\sqrt{3}\kappa\tau_{\gamma})$, and it can be shown that
$(x^{(3)},y^{(3)})$ is stable for $\alpha>-1$ and $\gamma >
\frac{\sqrt{3}\kappa\tau_{\gamma}}{1-\sqrt{3}\kappa\tau_g}$.

In the case of $\gamma=1$, the system has only two critical points,
the first one is $(x^{(*)},
y^{(*)})=\Big(\frac{\tau_g}{\tau_g+\tau_{\gamma}}, 0\Big)$, the
second one is still Eq.(20) with everywhere replaced by $\gamma=1$ .
At the first critical point, the eigenvalues are
$\lambda_1^{(*)}=-3\sqrt{3}\kappa(\tau_g+\tau_{\gamma})$ and
$\lambda_2^{(*)}=3(\alpha+1)(1-\sqrt{3}\kappa(\tau_g+\tau_{\gamma})$,
and so the point is unstable if $\alpha>-1$ and
$1-\sqrt{3}\kappa(\tau_g+\tau_{\gamma})>0$. It can be shown that the
second critical point is stable under the same assumptions. So we
obtain the same stable point in both cases. This critical point
corresponds to a GCG-dominated phase

\begin{equation}
\Omega_g=1-\frac{\sqrt{3}\kappa\tau_{\gamma}}{\gamma},
\Omega_{\gamma}=\frac{\sqrt{3}\kappa\tau_{\gamma}}{\gamma}
\end{equation}

\noindent and its equation of state is

\begin{equation}
w_g=-1+\frac{\sqrt{3}\gamma\kappa\tau_g}{\gamma-\sqrt{3}\kappa\tau_{\gamma}}
\end{equation}

\noindent In the $\tau_{\gamma}=\tau_g=0$ case, we have
$\Omega_g=1$ and $w_g=-1$ which is a late time de Sitter
attractor[21]. In the $\tau_{\gamma}=0$ and $\tau_g\neq 0$ case,
this analysis is consistent with the results of exact solution
(6).

\vspace{0.4cm} \noindent\textbf{4. Numerical analysis}
 \vspace{0.4cm}

Next, we study the above dynamical system numerically. For
definiteness, we choose the parameters to be $\gamma=1$ and
$\alpha=0.5$. The initial values of $x$ and $y$ are chosen as shown
in TABLE I and the results are contained in FIG.1-FIG.3. From FIG.1,
we can observe that all orbits tend to an attractor which
corresponds to $w_g=-\frac{91}{97}$. From FIG.2, we can observe that
for different initial values of $\rho_g$ and $p_g$, the equation of
state $w_g$ approaches the attractor $w_g=-\frac{91}{97}$ from
either $w_g>-1$ or $w_g<-1$. Irrespective of whether the initial
choice has $w_g>-1$ or $w_g<-1$, the equation of state will
eventually mimic that of quintessence in the viscous cosmology. It
is worth noting that if we choose the parameters so that the GCG
behaves as phantom, it is no longer possible to make it behave as
matter at an early epoch unless the viscosity coefficients are
anomalously large. However, in our setup of this paper, we have
included a linear barotropic fluid that could be used to mimic the
matter sector of our universe and thus GCG can be considered only as
dark energy. The evolution of the sound speed $c_s^2$ is shown in
FIG.3, where $c_s^2$ tends to $\frac{91}{194}$ for different initial
values of $\rho_g$ and $p_g$, and $\sqrt{3}\kappa\tau_{\gamma}=0.03,
\sqrt{3}\kappa\tau_g=0.06$.

The initial values of $x$ and $y$ are chosen as shown in TABLE II
and the results are contained in FIG.4. The different values of the
parameters $\tau_{\gamma}$ and $\tau_g$ are chosen as shown in TABLE
III and the results are contained in FIG.5.

\begin{center}
\begin{table}
\caption{The initial values of $x$ and $y$ in the plots FIG.1-FIG.3}
\begin{tabular}{ c  c c c c c c }
  \hline
  x & 0.14 & 0.15 & 0.16 & 0.17 & 0.18 & 0.18  \\
  \hline
  y & -0.19 & -0.18 & -0.17 & -0.16 & -0.15 & -0.14 \\
  \hline
\end{tabular}

\end{table}
\end{center}

\begin{figure}
\epsfig{file=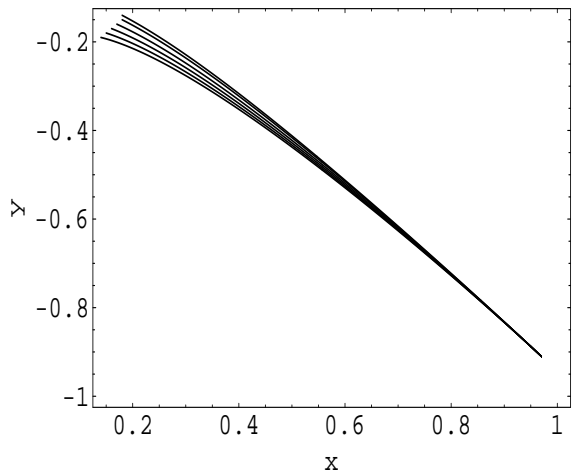,height=2.5in,width=3in} \caption{The phase
diagram of the viscous GCG system in terms of $x$ and $y$ for
different initial values of $x$ and $y$, and
$\sqrt{3}\kappa\tau_g=0.06, \sqrt{3}\kappa\tau_{\gamma}=0.03$.}
\end{figure}

\begin{figure}
\epsfig{file=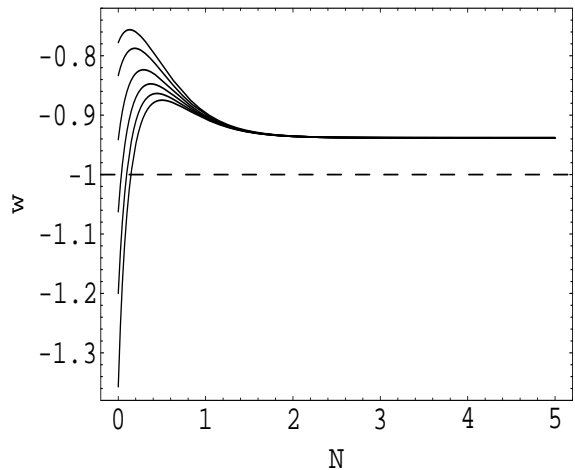,height=2.5in,width=3in} \caption{The evolution
of the equation of state of viscous GCG for different initial values
of $x$ and $y$, where we have taken $\sqrt{3}\kappa\tau_g=0.06,
\sqrt{3}\kappa\tau_{\gamma}=0.03$. The curves from bottom to top
correspond to the initial conditions specified in Table I from left
to right respectively.}
\end{figure}

\begin{figure}
\epsfig{file=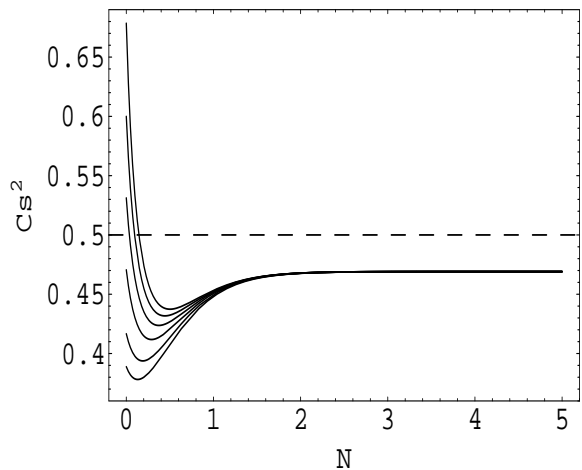,height=2.5in,width=3in} \caption{The evolution
of the sound speed  $c_s^2$ for different initial values of $x$ and
$y$ in a viscous cosmological model with $\sqrt{3}\kappa\tau_g=0.06,
\sqrt{3}\kappa\tau_{\gamma}=0.03$. The curves from top to bottom
correspond to the initial conditions specified in Table I from left
to right respectively.}
\end{figure}

\begin{center}

\begin{table}
\caption{The initial values of $x$ and $y$ in the plot FIG.4}
\begin{tabular}{ c  c c c c c }
  \hline
  x & 0.0140 & 0.0240 & 0.0340 & 0.0440 & 0.0540 \\
  \hline
  y & -0.0042 & -0.0036 & -0.0030 & -0.0024 & -0.0018 \\
  \hline
\end{tabular}

\end{table}
\end{center}

\begin{center}
\begin{table}
\caption{The parameter values of $\tau_g$ and $\tau_{\gamma}$ }
\begin{tabular}{ c  c c c c c c }
  \hline
  $\sqrt{3}\kappa\tau_g$ & 0.02 & 0.04 & 0.06 & 0.08 & 0.10 & 0.12  \\
  \hline
  $\sqrt{3}\kappa\tau_{\gamma}$ & 0.01 & 0.02 & 0.03 & 0.04 & 0.05 & 0.06 \\
  \hline
\end{tabular}

\end{table}
\end{center}

\begin{figure}
\epsfig{file=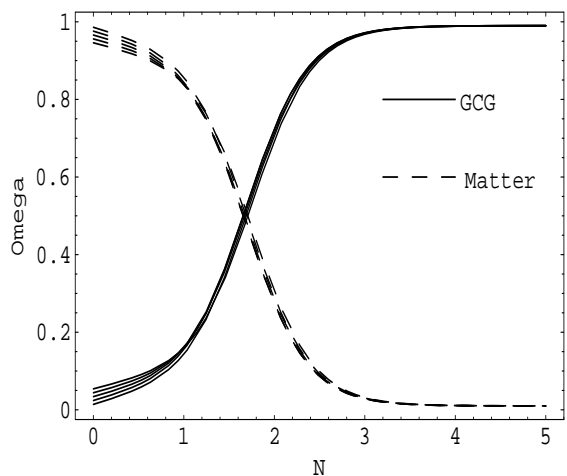,height=2.5in,width=3in} \caption{The evolution
of the cosmic density parameter for matter $\Omega_{\gamma}$ and
$\Omega_{g}$ respectively at different initial values of $x$ and
$y$, and$\sqrt{3}\kappa\tau_g=0.02,
\sqrt{3}\kappa\tau_{\gamma}=0.01$. }
\end{figure}

\begin{figure}
\epsfig{file=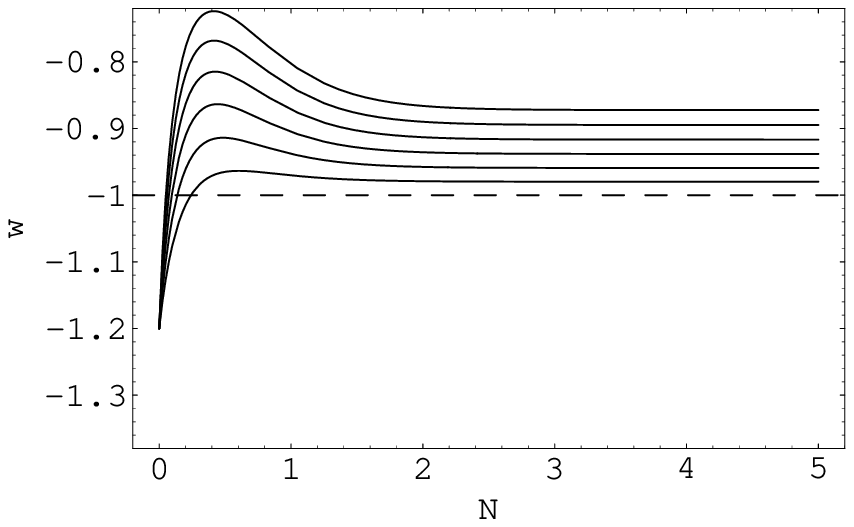,height=2.5in,width=3in} \caption{The evolution
of the equation of state of viscous GCG for different bulk viscosity
coefficients, where we have taken $x=0.15$ and $y=-0.18$. The curves
from bottom to top correspond to the parameter values specified in
Table 3 from left to right, respectively. }
\end{figure}

\vspace{0.4cm} \noindent\textbf{5. Conclusions}
 \vspace{0.4cm}

In the present paper, we have analyzed the dynamical evolution of
viscous GCG for different parameters and initial conditions. Our
results are as follows:

(1) We have shown that different initial values of $\rho_g$ and
$p_g$ will lead to different tracks ($w_g>-1$ and $w_g<-1$ ) for the
equation of state $w_g$ as it approaches the dynamical attractor
phase
$w_g=-1+\frac{\sqrt{3}\gamma\kappa\tau_g}{\gamma-\sqrt{3}\kappa\tau_{\gamma}}$.
That is to say, the comparison of theoretical values to
observational ones places a constraint on the bulk viscosity
coefficient. Recent astrophysical data indicate that the effective
equation of state parameter $w_{eff}$ lies in the interval:
$-1.38<w_{eff}<-0.82$ \cite{melchiorri}, so that we have the
constraint $\sqrt{3}\kappa(\frac {50}{9}\tau_g+\tau_{\gamma})<1$.

(2) We can also use the requirement that $0<c_s^2<1$ in the GCG to
place the constraint
$0<\alpha<1+\frac{\sqrt{3}\kappa\gamma\tau_g}{\gamma-\sqrt{3}
\kappa\tau_{\gamma}-\sqrt{3}\kappa\gamma\tau_g}$ on the parameter
$\alpha$ in the model .

(3)The equation of state $w_g$ can cross the boundary $w_g=-1$.

(4) From the point of view of the dynamics, the bulk viscosity
coefficient should satisfy
$\gamma>\frac{\sqrt{3}\kappa\tau_{\gamma}}{1-\sqrt{3}\kappa\tau_g}$.

(5)In the viscous model, the ratio of cosmic density parameters
$\frac{\Omega_{\gamma}}{\Omega_g}$ approaches the constant
$\frac{\sqrt{3}\kappa\tau_{\gamma}}{\gamma-\sqrt{3}\kappa\tau_{\gamma}}$.
If we suppose the present universe is in the epoch of
$\frac{\Omega_{\gamma}}{\Omega_g}$ approximating to a constant, we
can obtain the constraints $\sqrt{3}\kappa\tau_{\gamma}=0.3\gamma$
and $\sqrt{3}\kappa\tau_g<0.7$.

Finally, it is worth noting that as a phenomenological model of the
evolution of the late universe, it  is reasonable for our model to
leave the causal viscosity theory out of account.

\vspace{0.8cm} \noindent ACKNOWLEDGEMENT: This work is supported
by National Nature Science Foundation of China under Grant No.
10473007 and Shanghai Rising-Star program under Grant
No.02QA14033.


\begin{thebibliography}{99}
\bibitem {melchiorri}A. Melchiorri , L. Mersini-Houghton , C. J. Odmann
and M. Trodden \textit{Phys. Rev.}, \textbf{D68}, 043509 (2003).
\bibitem {caldwell} R. R. Caldwell,  M. Kamionkowski and N.
N. Weinberg \textit{Phys. Rev. Lett.}, \textbf{91}, 071301 (2003);
H. Stefancic \textit{Phys. Rev.}, \textbf{D71}, 124036 (2005); J. D.
Barrow \textit{Class. Quantum Grav.}, \textbf{21}, L79 (2004); J. D.
Barrow and C. G. Tsagas  \textit{Class. Quantum Grav.}, \textbf{22},
1563 (2005); S. Nojiri,  S. D. Odintsov and  S. Tsujikawa
\textit{Phys. Rev.} \textbf{D71}, 063004 (2005).
\bibitem{mcinnes} B. McInnes  \textit{JHEP}, \textbf{0208}, 029 (2002).
\bibitem {Peebles}
 P. J. E. Peebles and B. Ratra \textit{Rev. Mod. Phys.}, \textbf{75}, 599 (2003);
T. Padmanabhan \textit{Phys. Rep.}, \textbf{380}, 235 (2003).
\bibitem{li}X. Z. Li , J. G. Hao and D. J. Liu \textit{Class. Quantum
Grav.}, \textbf{19}, 6049 (2002); X. Z. Li and J. G. Hao
\textit{Phys.Rev.}, \textbf{D69}, 107303 (2004).
\bibitem {well} R. R. Caldwell \textit{Phys. Lett. }, \textbf{B545}, 23 (2002).
\bibitem{hao3}J. G. Hao and X. Z. Li \textit{Phys. Rev.}
\textbf{D68}, 043501 (2003); D. J. Liu and X. Z. Li
\textit{Phys.Rev.}, \textbf{D68}, 067301 (2003).
\bibitem{armendariz} C. Armendariz-Picon, V. Mukhanov and P. J. Steinhardt \textit{Phys. Rev. }, \textbf{D63}, 103510 (2001).
\bibitem{GCG}M. Makler, S. Q. Oliveira, and I. Waga \textit{Phys.
Lett.}, \textbf{B555}, 1 (2003); M. C. Bento, O. Bertolami and A. A.
Sen \textit{Phys. Rev.}, \textbf{D67}, 063003 (2003); D. J. Liu and
X. Z. Li \textit{Chin. Phys. Lett. }, \textbf{22}, 1600 (2005).
\bibitem{chgas}A. Kamenshchik, U. Moschella and V. Pasquier \textit{Phys. Lett.},
 \textbf{B511}, 265 (2001);
N. Bili¡äc, G. B. Tupper and R. D. Viollier \textit{Phys. Lett.},
\textbf{B535}, 17 (2002).
\bibitem {hao} J. G. Hao and X. Z. Li \textit{Phys. Lett.},
\textbf{B606}, 7 (2005).
\bibitem{barrow} J. D. Barrow \textit{Phys. Lett.}, \textbf{B180}, 335 (1987);
I. Brevik , S. D. Odintsov \textit{Phys. Rev. }, \textbf{D65},
067302 (2002); D. J. Liu and X. Z. Li \textit{Phys. Lett.},
\textbf{B611}, 8 (2005).
\bibitem{eckart}C. Eckart \textit{Phys. Rev.}, \textbf{58}, 919 (1940).
\bibitem{landau} L. D. Landau and E. M. Lifshitz \textit{Fluid
Mechanics} (Butterworth Heinemann,1987).
\bibitem{israel} W. Israel \textit{Ann. Phys.}, \textbf{100},
310 (1976).
\bibitem{stewart} W. Israel and J. M. Stewart \textit{Phys.
Lett.}, \textbf{A58}, 213 (1976); W. A. Miscock and J. Salmomson
 \textit{Phys. Rev.}, \textbf{D43}, 3249 (1991); R. Maartens
\textit{Class. Quantum Grav.}, \textbf{12}, 1455 (1995).
\bibitem{harko} T. Harko and M. K. Mak \textit{Class. Quantum
Grav.}, \textbf{20}, 407 (2003).
\bibitem{brevik} I. Brevik and O. Gorbunova [gr-qc/0504001]
\bibitem{zimdahl} W. Zimdahl and D. Pav\'{o}n \textit{Phys. Rev.},
\textbf{D61}, 108301 (2000).
\bibitem{gron} {\O}. Gr{\o}n  \textit{Astrophys. Space Sci.},
\textbf{173}, 191 (1990).
\bibitem{reis} R. R. R. Reis, M. Makler, S. Q. Oliveira, I. Waga
 \textit{Phys. Rev.}, \textbf{D69}, 101301(2004).
\bibitem{sandvik} H. Sandvik, M. Tegmark, M. Zaldarriaga, I. Waga
 \textit{Phys. Rev. }, \textbf{D69}, 123524 (2004).
\bibitem {haonew} J. G. Hao and X. Z. Li \textit{Phys. Rev.},

\textbf{D70}, 083514 (2004).
\bibitem{li2}X. Z. Li, Y. B. Zhao and C. B. Sun \textit{Class. Quantum
Grav.},
 \textbf{22},3759 (2005).



\end{thebibliography}
\end{document}